\documentclass[aps,prl,floatfix,amsmath,amssymb,nofootinbib,twocolumn,groupedaddress]{revtex4}
\usepackage[dvips]{graphicx}
\usepackage{graphicx}
\usepackage{amsmath}
\usepackage{amssymb}
\usepackage{amsfonts}
\usepackage{upgreek}
\usepackage{txfonts}
\usepackage{color}
\usepackage{yfonts}
\usepackage{latexsym}
\usepackage{amsmath}
\usepackage{bm}
\newcommand{\beq}{\begin{equation}}
\newcommand{\eeq}{\end{equation}}
\newcommand{\ba}{\begin{array}{ccc}}
\newcommand{\ea}{\end{array}}
\newcommand{\nn}{\nonumber \\}
\newcommand{\bx}{{\bm x}}
\newcommand{\bk}{{\bm k}}
\newcommand{\bp}{{\bm p}}
\newcommand{\bq}{{\bm q}}
\newcommand{\bl}{{\bm l}}
\def\bea{\begin{eqnarray}}
\def\eea{\end{eqnarray}}
\begin{document}

\title{Vector boson excitations near deconfined quantum critical points}

\author{Yejin Huh}
\email{yejinhuh@fas.harvard.edu}
\author{Philipp Strack}
\author{Subir Sachdev}
\affiliation{Department of Physics, Harvard University, Cambridge MA 02138}

\date{\today}

\begin{abstract}
We show that the N\'eel states of two-dimensional antiferromagnets have low energy vector boson excitations in the vicinity
of deconfined quantum critical points. We compute the universal damping of these excitations arising from spin-wave  emission.
Detection of such a vector boson will demonstrate the existence of emergent topological gauge excitations in a quantum spin system. 
\end{abstract}

\maketitle

Quantum spin systems are expected to display ground states with novel 
fractionalized and topological gauge excitations which have no analogs in the band insulators of Bloch theory \cite{Balents}.
The theoretical arguments for the existence of such excitations are convincing, but, so far, the excitations have
not been detected unambiguously in any experiment, or even in numerical studies of semi-realistic model systems
in two spatial dimensions. In experiments, the best candidate so far is the kagome antiferromagnet, and its neutron
scattering spectrum displays strong evidence for fractionalization \cite{Han}, but the specific excitations have not been identified.
In numerics, there has been positive evidence for exotic physics in the topological entanglement entropy of the 
kagome antiferromagnet \cite{Balents2,Schollwoeck}, 
but this does not directly identify the excitation spectrum. Sandvik \cite{sandvik} has obtained 
convincing evidence of an emergent U(1) symmetry near the quantum critical point of a spin system, and this
is strong, but indirect, evidence of an emergent, topological gauge excitation of a deconfined quantum critical point \cite{science04};
however, this does not yield any information on the excitation spectrum of the gauge boson.

In this paper we propose that the antiferromagnetically ordered ({\em i.e.\/} N\'eel) phase of a
quantum spin system near a deconfined quantum critical point has an emergent 
vector boson excitation which should
be detectable in numerical studies, and possibly eventually in experiments. This vector boson is the analog
of the $W$ and $Z$ vector bosons of the standard model of particle physics, and is similarly a characteristic signature of the gauge
structure of the underlying theory. In two spatial dimensions, the vector boson is universally damped by emissions of multiple 
spin-wave ({\em i.e.\/} Goldstone boson) excitations of the N\'eel phase, and the present paper will provide a quantitative
computation of this damping. The N\'eel phase also has a universally damped Higgs excitation \cite{podolsky1}, 
and this mode has recently been detected
in experiments \cite{endres} and numerics \cite{podolsky2,chen13} near a conventional quantum critical point. We argue here that similar methods
can allow positive identification of a vector boson excitation in the N\'eel phase near a deconfined quantum critical point.
Note that, while the Higgs mode is present for both conventional and deconfined critical points, the vector boson mode appears only in the latter case.

We note that there is a debate \cite{anders1,nahum,boris,white,ganesh,damle,block13,bergerhoff96,lorenz} in the literature of whether the gauge theory of the antiferromagnet ultimately
describes a second-order quantum phase transition in antiferromagnets with a global SU(2) spin rotation symmetry. 
Our method of detecting the vector boson sidesteps this delicate issue,
because the vector boson should exist even if the transition out of the N\'eel phase is weakly first-order. Its observation 
would be a direct signature that the theory of deconfined criticality with emergent gauge excitations describes the spectrum
of the antiferromagnet at low energies.

For quantum antiferromagents with SU($N$) global symmetry, the deconfined critical theory \cite{rsprb,science04} is the CP$^{N-1}$ field theory
of relativistic complex scalars (`spinons') $z_\alpha$ ($\alpha = 1 \ldots N$) minimally coupled to a U(1) gauge field $A_\mu$
with partition function 
%
$Z = \int \mathcal{D} z_\alpha \mathcal{D} \lambda \mathcal{D}  A_\mu e^{-\mathcal{S}}$ with
\begin{align}
\mathcal{S} &= \int_\bx \left[ \frac{N}{g} |(\partial_\mu - i A_\mu ) z_\alpha|^2 + i \lambda (|z_\alpha|^2 - 1)
 \right]\; ,
\label{eq:Z}
\end{align}
where the integration is over 2+1 dimensional space and (imaginary) time, and $\lambda$ is a Lagrange multiplier which implements the constraint
$\sum_{\alpha=1}^N |z_\alpha|^2=1$.
For $N=2$, the N\'eel order parameter, $\mathbf{n}(\mathbf{x})$, is related to the spinons via
\begin{align}
\mathbf{n} = z^\ast_\alpha\, \boldsymbol{\sigma}_{\alpha\beta} \, z_\beta\;,
\label{eq:CP_1}
\end{align}
where $\boldsymbol{\sigma}$ is a vector of Pauli matrices.

We are interested here in the spectrum of $A_\mu$ excitations in the N\'eel phase at zero temperature, which appears for $g< g_c$
with $\left\langle \mathbf{n}(\mathbf{x}) \right\rangle \neq 0$. We will detect this spectrum via correlations of the 
staggered vector spin chirality
\beq
B_\mu = \epsilon_{\mu\nu\lambda} \partial_\nu A_\lambda = {\textstyle \frac{1}{4}} \epsilon_{\mu\nu\lambda} \mathbf{n} \cdot \left( \partial_\nu \mathbf{n} \times
\partial_\lambda \mathbf{n} \right) ; \label{chirality}
\eeq
the last term specifies how $B_\mu$ can be related to the operators of the underlying antiferromagnet,
and identifies it as the Skyrmion current: the spatial integral of its temporal component $B_t$ is the Skyrmion number of the
texture of the N\'eel order parameter underlining the topological nature of the vector boson.
Indeed, correlations of 
$\epsilon_{\mu\nu\lambda} \mathbf{n} \cdot \left( \partial_\nu \mathbf{n} \times
\partial_\lambda \mathbf{n} \right)$ were measured recently by Fritz {\em et al.} \cite{fritz} in quantum Monte Carlo.
They can also be measured in Raman scattering \cite{ss,nl} if the light couples preferentially to one sublattice
of the antiferromagnet. 

In the vicinity of a deconfined critical point, we show that the existence of an emergent gauge field implies that the correlations
of $B_\mu$ obey the universal scaling form
\beq
\left\langle B_\mu (-p) B_\nu (p) \right\rangle  = \left( \delta_{\mu\nu} - \frac{p_\mu p_\nu}{p^2} \right)\,  \rho_s \, \mathcal{F} \left( {p}/{\rho_s} \right),
\label{transverse}
\eeq
where $p$ is a Euclidean 3-momentum, $\rho_s$ is the `spin stiffness' (or `helicity modulus') of the N\'eel phase, 
and $\mathcal{F}$ is a completely universal scaling function (including its overall scale). The stiffness vanishes as $\rho_s \sim (g-g_c)^{\nu}$ upon
approaching the deconfined critical point, where $\nu$ is the correlation length exponent. We also note that $\rho_s$ is analogous to the $f^2$ constant
of the chiral Lagriangian of particle physics. Here, $\rho_s$ serves as the low energy scale controlling the excitations of the N\'eel phase.

The correlator in Eq.~(\ref{transverse}) is purely transverse; lattice models will also have a longitudinal component, but this is suppressed near the deconfined critical point. 
In contrast, for a conventional critical point, the longitudinal and transverse components both vanish
rapidly with the same large power $ \sim \rho_s^{2 \Delta_B - 3}$ (and non-universal prefactor), where \cite{fritz} $\Delta_B \sim 3.4$ 
is the scaling dimension of the staggered spin chirality (compare to the linear $\sim \rho_s$ power in Eq.~(\ref{transverse})).

\begin{figure}[]
\includegraphics[width=40mm]{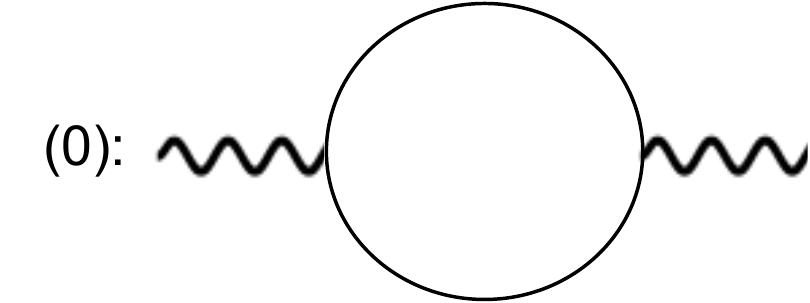}
\caption{$N\rightarrow\infty$ contribution to the gauge propagator that generates purely dissipative dynamics by 
decaying into two complex Goldstone bosons leading to Eq.~(\ref{eq:f_infty}).
Wiggly line refers to the gauge field and the solid line the $N$ complex Goldstone bosons $\pi$.}
\label{fig:n_infty}
\end{figure}

\begin{figure}[t]
\includegraphics[width=90mm]{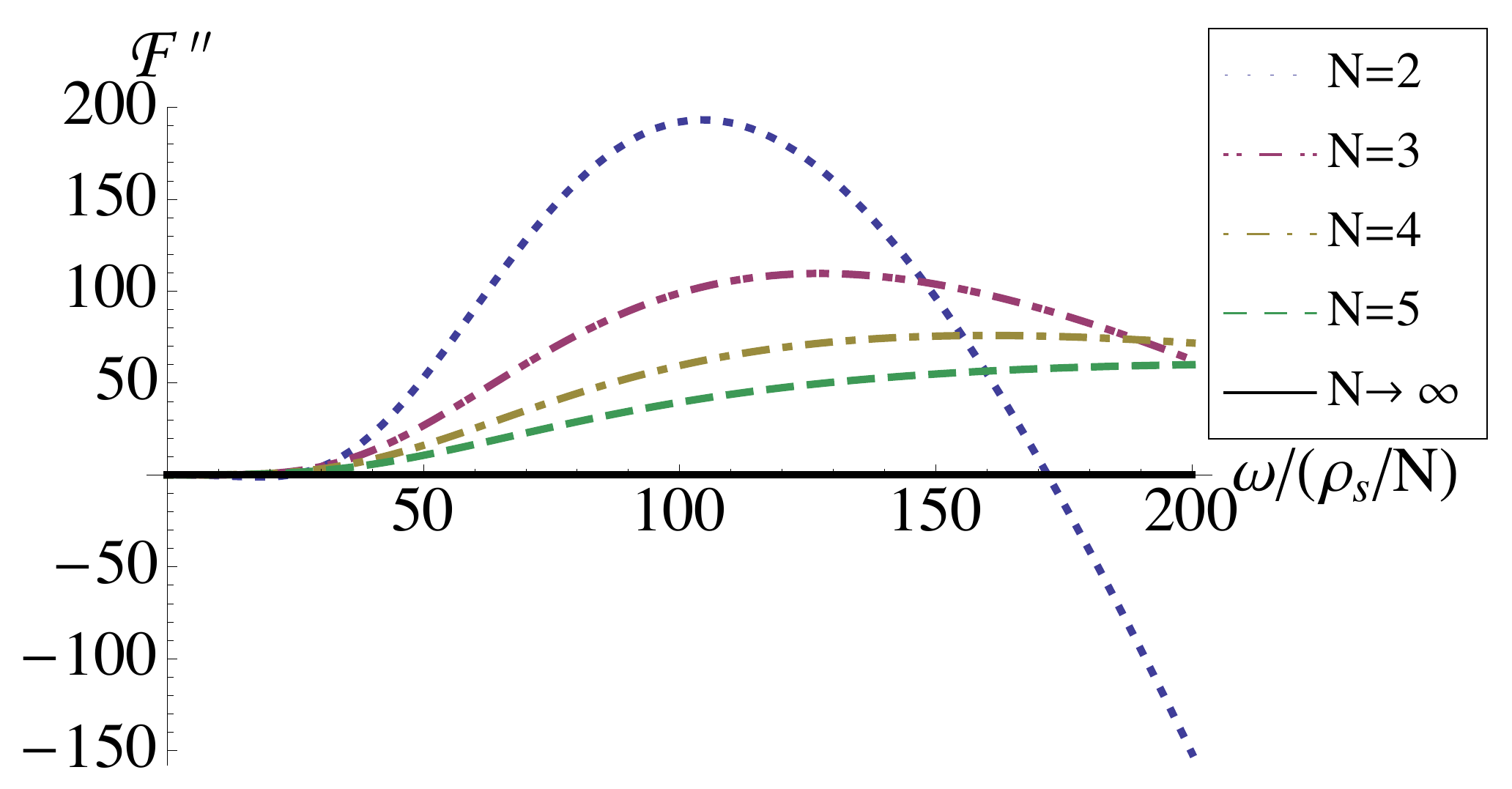}
\caption{(Color online) Spectrum of the vector boson as obtained from Eqs.~(\ref{eq:DH}, \ref{eq:im_F}) 
as a function of real frequencies $\omega$ at zero spatial momenta 
displaying the vector boson resonance (in a quantum spin system near a deconfined quantum critical point) as
$N$ approaches physically relevant 
values toward $N=2$ (blue, dotted curve); the change in sign for large $\omega$ at $N=2$ is due to the breakdown of the $1/N$ expansion, as in 
Ref.~\cite{podolsky1}. 
Frequencies are given 
in units of the spin stiffness $\rho_s$, Eq.~(\ref{rhos}) plotted for $\mathcal{C}=0$. Shifting $\mathcal{C}$ away from $0$ does not qualitatively change the curves.}
\label{fig:imphi}
\end{figure}

This paper will obtain numerous properties of the scaling function 
$\mathcal{F}$ for the deconfined critical point in the $1/N$ expansion.
It is useful to first present the form of these gauge-field correlators in the large $N$ limit. After continuing the result to Lorentzian frequencies 
at zero spatial momentum
($p \rightarrow - i \omega$) we have
\beq
\rho_s \mathcal{F} \left( \omega /\rho_s \right) = \frac{1}{N} \frac{ - 16 \omega^2}{(- i \omega + 64 \rho_s /N)};
\label{eq:f_infty}
\eeq
note that the large $N$ limit is taken with $\rho_s/N$ fixed. This pole in the lower-half of the complex frequency plane
at $\omega = - i 64 \rho_s/N$, arising from the rapid decay due to spin-wave emission (cf. Fig.~\ref{fig:n_infty}), represents the overdamped vector boson excitation.
The large $N$ status of the vector boson is therefore similar to that of the Higgs boson in the N\'eel phase \cite{podolsky1}. 
And just as was the case for the Higgs boson, we will find here that $1/N$ corrections lead to a non-zero real part in the position of the vector boson pole (see Eq.~(\ref{eq:pole}) below), so that at small $N$ we expect that the real and imaginary parts are both of order $\rho_s$. Moreover, the imaginary part of $\mathcal{F}$ on the real frequency axis, shown in Fig.~\ref{fig:imphi}, displays a vector boson resonance 
for the physically relevant $N$.

The tensor structure of the vector boson correlator, along with spontaneously broken gauge symmetry in a Higgs phase, represented
significant technical obstacles, making this computation more challenging than previous computations of critical properties in the $1/N$ expansion. 

We now outline our computation.
First, without loss of generality, we choose 
the $z_\alpha$ condensate along the flavor index $\alpha=1$ direction, and parameterize
$z(\bx)=\left( \sigma(\bx)e^{i\omega(\bx)},
\pi_1(\bx),
\pi_2(\bx),
\dotsc,
\pi_{N-1}(\bx) \right)$
where the $\pi_i$-fields are complex-valued and $\sigma(\bx)$ and $\omega(\bx)$ are real-valued.
It is convenient to use a radial 
coordinate system for the first flavor component so that the $N$-component. 
As a consequence of this 
coordinate transformation, the measure of the functional integral for the first flavor component at each point $\bx$ picks up a 
Jacobian, 
det $J=\sigma$ \cite{munster}.
In unitary gauge, 
the (redundant) local gauge transformation function is chosen as the phase variable of the first flavor $\omega(\bx)$ 
\cite{appelquist73,munster}. Then, as usual, the Goldstone boson of the first flavor is ``eaten up'' and the 
action does not depend on $\omega(\bx)$ anymore. 
In the large $N$ limit, we find a saddle point with $\sigma = \sqrt{N} \sigma_0$ with
$\sigma_0^2 = {1}/{g} - \int_\bp {1}/{p^2}$.
For the amplitude fluctuations around this condensate, we shift $\sigma \rightarrow \sqrt{N} \sigma_0 + \sigma$.
It is crucial to perform the shift in $\sigma$ also for the Jacobian, and re-exponentiate
it as a propagator $\langle \bar{c} c \rangle$ of fermionic ghost fields $\bar{c}$, $c$. Note that 
the inclusion of ghost tadpole diagrams is crucial to ensure that the mass of the Goldstone bosons 
($\pi_i$'s) stays identically zero \cite{huh13}.
%
%
%
\begin{figure}[]
\includegraphics[width=40mm]{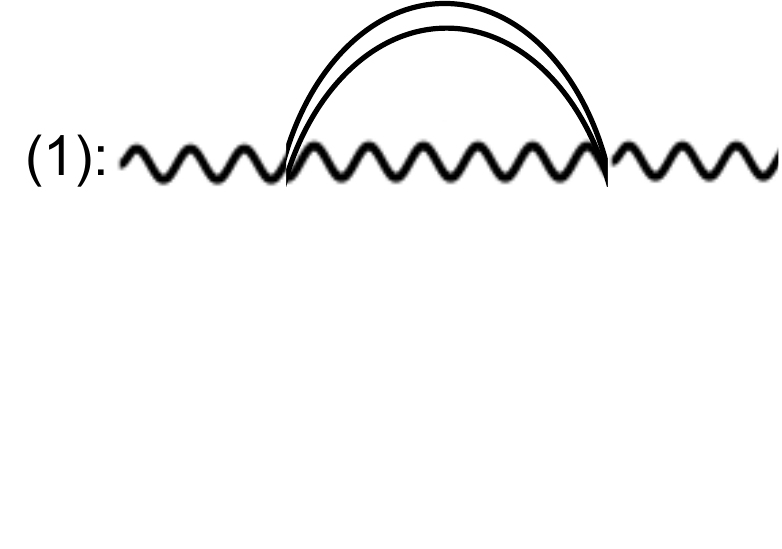}\\[-15mm]
------------------------------------------------------------------------\\[2mm]
\includegraphics[width=90mm]{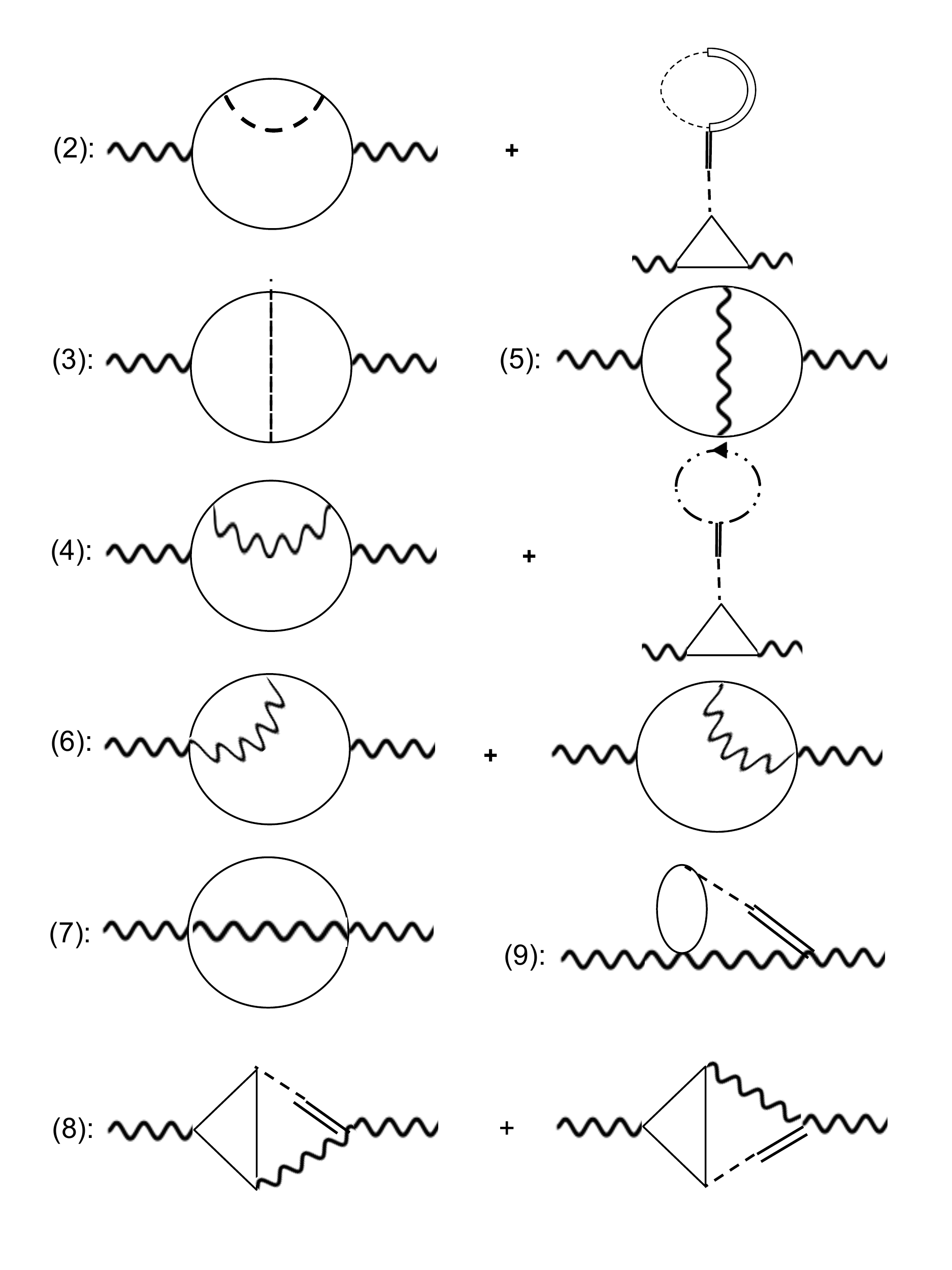}\\[0mm]
------------------------------------------------------------------------\\[7mm]
\includegraphics[width=85mm]{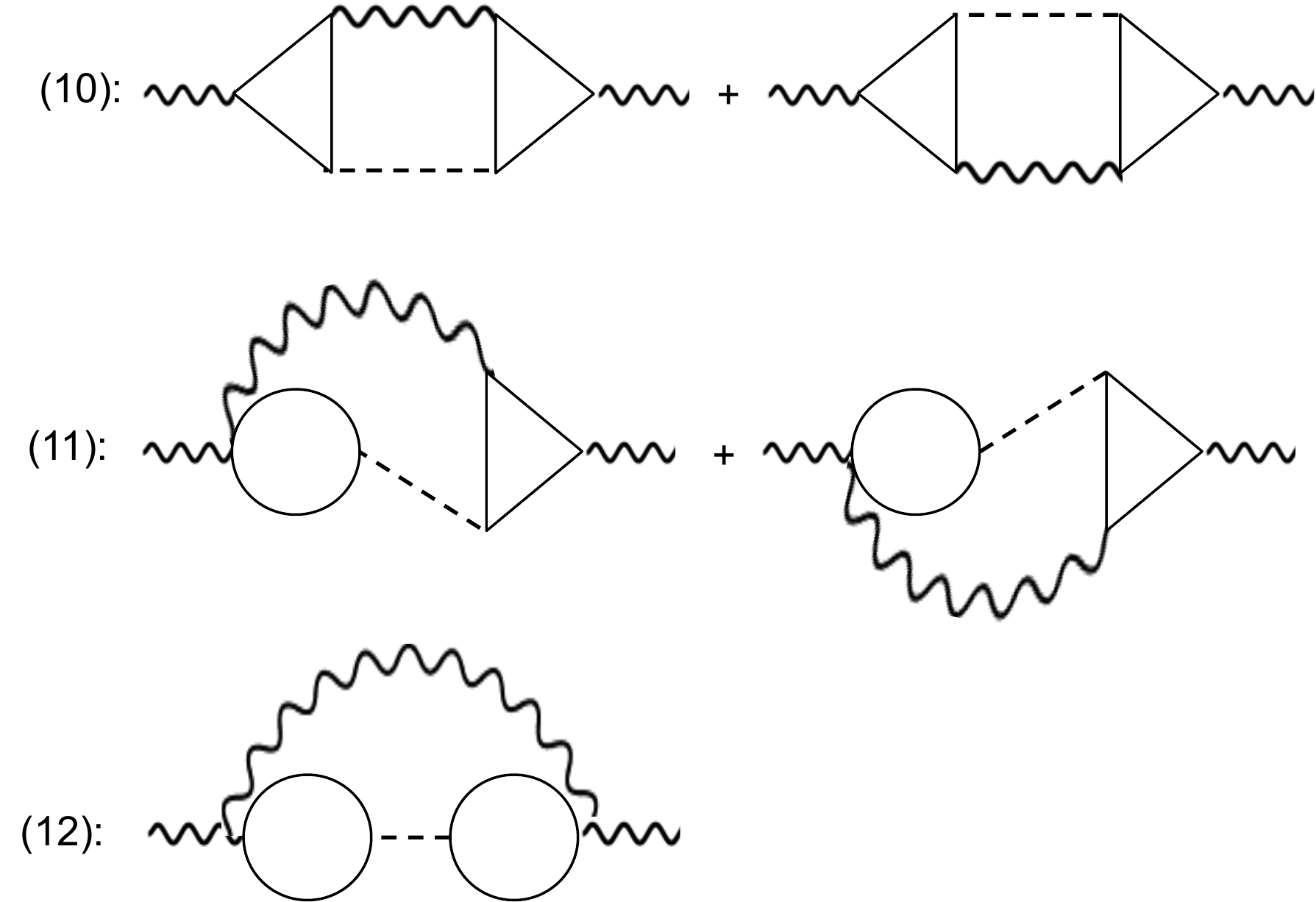}
\caption{1-, 2- and 3-loop diagrams that renormalize the gauge propagator to order $1/N$ after performing the 
flavor trace. Conventions as in Fig.~\ref{fig:n_infty}, the double line represents the longitudinal $\sigma$ field, the dotted line represents the Lagrange multiplier $\lambda$ field, and the dot-dashed line the fermionic ghost field $c$. The analytic expressions including factors are given in the Supplemental Material.}
\label{fig:2loop}
\end{figure}
%
%

We now perform the large-$N$ expansion \cite{polyakov_book}. The Feynman rules for the various 
vertices can be obtained from the original action Eq.~(\ref{eq:Z}) after the previously mentioned substitutions of the
$z$-fields. We then integrate out the 
complex Goldstone fields $\pi$, $\bar{\pi}$ and expand the still dynamical determinant 
to quadratic order in the fields $A_\mu$, $\sigma$ and $\lambda$ \cite{kaul08}. This 
yields the form of the $N\rightarrow\infty$ propagators 
$\langle\sigma\sigma\rangle, \langle\lambda\lambda\rangle, \langle\sigma\lambda\rangle$ 
and $\langle A_\mu A_\nu\rangle$. The gauge field and Lagrange multiplier become 
dynamical in this way.
We then evaluate Wick's theorem keeping all contributions to order $1/N$ after performing 
the internal flavor trace. Our result for the self-energy 
of the gauge bosons are the 12 diagrammatic contractions shown in Fig.~\ref{fig:2loop}, denoted 
by 
\begin{align}
\Sigma_{\mu\nu} (p) = \sum_{i=1}^{12} a_i \Sigma^{(i)}_{\mu\nu} (p)\;,
\end{align}
where we have separated equivalent contractions into numerical factors $a_i$. 
The explicit expressions are given in the Supplemental Material. In the past, a reliable evaluation of 
such tensor-valued Feynman diagrams in momentum space for vectorial correlation functions has been 
almost intractable. Already for the much simpler case of the (conformally invariant) fixed point of the $O(N)$ 
vector model, a computation of vectorial correlation functions to $1/N$ is an intricate matter, necessitating the 
use of conformal field theory methods in real space \cite{osborn94,petkou95,petkou96,erdmenger97}. In momentum space, 
Cha {\it et al.} \cite{cha91} have succeeded to compute the current-current correlator of the $O(N)$ model to $1/N$ 
but needed to supplement their calculation with results from other computations; and their method seems hard to 
generalize to more complicated situations. 

From a technical viewpoint, the enabling achievement of this 
paper is the reliable, numerically verifiable computation of tensor-valued momentum integral of multi-loop 
diagrams using our algorithm Tensoria \cite{huh13} (see Refs.~\onlinecite{bzowski12,suvrat13} for 
an application of similar methods to three-point functions of conformal field theories).

Including the 12 contractions of Fig.~\ref{fig:2loop}, we write the renormalized gauge propagator $D_{\mu\nu}(p)=N \langle A_\mu(-p) A_\nu(p) \rangle$ as
\beq
\left[ D_{\mu\nu} (p) \right]^{-1}
= \left(1 - \frac{1}{N} \right) \frac{p}{16} \left( \delta_{\mu \nu} - \frac{p_\mu p_\nu}{p^2} \right) + 2 \sigma_0^2
\delta_{\mu\nu} - \Sigma_{\mu\nu} (p) \;,
\label{eq:D}
\eeq
where the terms not proportional to $1/N$ are in fact the $N\rightarrow \infty$ contribution shown in Fig.~\ref{fig:n_infty} that 
give the gauge boson its dynamics in the first place. 

Before describing the $1/N$-corrections further, we note here that our calculations passed several consistency checks.
First of all, we have checked by explicit computation that all diagrams that, individually, would generate a mass for the 
Goldstone bosons, instead cancel with each other. Moreover, we have computed the correlation length exponent 
$\nu = 1 - {48}/{(N\pi^2)}$ in agreement with previous work \cite{halperin74,irkhin96} by summing the prefactors of all 
logarithmic singularities $\sim \log [\Lambda/\sigma_0] \sigma_0^2$ 
(where $\Lambda$ is a momentum cutoff) appearing in the 12 self-energy diagrams of Fig.~\ref{fig:2loop}. 
Finally, we computed the gauge field propagator and the current-current 
correlator also at the critical point (for $\sigma_0=0$) \cite{huh13} and showed in particular that all (logarithmic and other) 
singularities as a function of momentum $p$ cancel with each other. Thereby, we demonstrated that, as expected \cite{gross75,franz03}, these conserved quantities do not pick up anomalous dimensions (beyond the $N\rightarrow\infty$ renormalization which is essentially a dimensional effect), and that they fulfill the expected Ward identities between 
self-energy and vertex corrections.

We now present our main quantitative results for the quantum dynamics of the vector gauge boson in the vicinity of 
the critical point. As announced earlier, we will
compute the position of the pole of the gauge-field propagator on the real frequency axis at zero spatial momentum, 
and the spectral function which can also be extracted in numerical simulations as well as potentially in experiments. 
In the large $N$ limit, the pole of the gauge-field propagator (zero in the transverse component of Eq.~(\ref{eq:D})) is located at $-32\sigma_0^2$, or $\omega=-32\sigma_0^2i$ in frequency with the Lorentzian time signature, corresponding to an
overdamped mode of purely dissipative character. This is just the well known consequence of the decay of the gauge boson into 
the ``particle and anti-particle'' continuum of Goldstone bosons (as per Fig.~\ref{fig:n_infty}) with the scale $32\sigma_0^2$ provided by the Higgs mechanism.

However, after accounting for the self-energy induced shifts of the pole position to $1/N$, the quantum dynamics of the gauge field 
changes qualitatively and gains an oscillatory response characteristic of a finite lifetime excitation similar to the scattering 
resonances of the W- and Z boson in particle physics.
To determine the shift, the $\Sigma_{\mu\nu}(p)$ integral is calculated at $p = - 32 \sigma_0^2$ numerically, as described in Ref.~\onlinecite{podolsky1}, by analytically continuing the contour integral. We take $k \rightarrow ke^{i \theta}$ and $p \rightarrow pe^{i \theta}$, then take the limit of $\theta \searrow -\pi$, rotating the contour around the lower half plane. The 
$\theta$ variation is done gradually, 
to ensure that we do not cross any poles. For a few of the diagrams, the integral can be done analytically by putting $p$ and $k$ right below the negative real axis, and the analytic and numerical results agree.

Finally, the universal structure of the pole position is revealed by writing $\sigma_0^2$ in terms of the 
spin stiffness $\rho_s$; their relationship is easily computed at $N=\infty$, and using the scaling law $\rho_s \sim (g-g_c)^{\nu}$ and the value
of $\nu$ quoted above, we obtain at order $1/N$
\beq
\rho_s = \frac{N \sigma_0^2}{2} \left( 1+ \frac{48}{N \pi^2} \ln \left( \frac{\Lambda}{32 \sigma_0^2} \right) + \frac{\mathcal{C}}{N} \right),
\label{rhos}
\eeq
where $\mathcal{C}$ is a constant of order unity.
Then, in terms of Lorentzian frequencies $\omega$, the pole is at
\beq
\frac{\omega_{\rm pole}}{\rho_s/N} = - 64i + \frac{32}{N}\left[ \, 7.319 - (11.191 - 2 \mathcal{C}) i \, \right] +   O(1/N^2)\;.
\label{eq:pole}
\eeq
In order to calculate the vector boson spectral function, we expand the renormalized propagator, 
Eq.~(\ref{eq:D}), to order $1/N$ by $D_{\mu\nu}(p)= D_{\mu\nu}^0(p)+ D_{\mu\nu}^0(p)\Sigma^{1/N}_{\mu\nu}(p) D_{\mu\nu}^0(p)$, and write 
\beq
D_{\mu\nu}(p)= H_T(p) \left( \delta_{\mu \nu} - \frac{p_\mu p_\nu}{p^2} \right) + H_L(p)  \frac{p_\mu p_\nu}{p^2} .
\label{eq:DH}
\eeq
The transverse part determines the response in Eq.~(\ref{transverse}), $\rho_s \mathcal{F} (p/\rho_s)=(p^2/N) H_T (p)$, and 
its imaginary part
\begin{align}
\mathcal{F}''(\omega/\rho_s) \equiv \frac{1}{\rho_s} \text{Im}\left[ \frac{p^2}{N} H_T(p)\Big|_{p\rightarrow - i\omega}\right]
\label{eq:im_F}
\end{align}
determines the spectral properties of the vector boson. The Wick rotation of $H_T$ to real frequencies 
has to be done numerically similarly to the procedure described above Eq.~(\ref{eq:pole}) except that now 
we rotate directly onto the imaginary axis at $\theta \searrow-\pi/2$. 

As can be observed from Fig.~\ref{fig:imphi}, the vector boson spectrum  
at $N\rightarrow\infty$ displays a suppressed, broad continuum indicative of the dissipative nature of the vector boson at that order. 
For the physically relevant smaller values of $N$, 
a peak-like structure emerges, which becomes progressively better defined upon approaching the case $N=2$ 
for deconfined quantum magnets. This indicates an enhanced lifetime of the vector boson. In real-time experiments, 
the vector boson response is that of a damped oscillator. As with the Higgs mode close to the superfluid-to-Mott insulator 
quantum phase transition \cite{podolsky1, chen13, endres}, we expect this emergent vector boson ``resonance''  for smaller $N$ 
to be observable in direct 
numerical simulations of $S\!U(N)$ 
quantum spin models \cite{block13,kaul12} using the observables of Ref.~\cite{fritz}, and potentially also in experiments. 

We acknowledge helpful discussions with Debanjan Chowdhury, Fabian Grusdt, Matthias Punk and Julian Sonner. 
This research was supported by the DFG under grant Str 1176/1-1, by the NSF under Grant DMR-1103860, 
by the John Templeton foundation, by the Center for Ultracold Atoms (CUA) and by the Multidisciplinary University Research Initiative (MURI). 
This research was also supported in part by Perimeter Institute for
Theoretical Physics; research at Perimeter Institute is supported by the
Government of Canada through Industry Canada and by the Province of
Ontario through the Ministry of Research and Innovation.

\clearpage

\begin{widetext}
\section*{Supplemental Material}
The self energy correction at order $1/N$ is a sum over 12 diagrams
$\Sigma_{\mu\nu} (p) = \sum_{i=1}^{12} a_i \Sigma^{(i)}_{\mu\nu} (p)$, where the $a_i$ are symmetry and 
multiplicity factors for each diagram not directly contained in the vertices and propagator Feynman rules.
Before the momentum integrations, the expressions for each of the diagrams are
\begin{alignat}{2}
\Sigma^{(1)}_{\mu\nu} (p) &= \frac{32 \sigma_0^2}{N} \int_\bq \left( \delta_{\mu\nu} 
  + \frac{q_\mu q_\nu}{32 q \sigma_0^2} \right)\frac{1}{(q + 32 \sigma_0^2)|\bp + \bq| (|\bp + \bq| + 16  \sigma_0^2)} 
  & a_1=4\nn
\Sigma^{(2)}_{\mu\nu} (p) &= -\frac{8}{N} \int_\bq \int_\bk 
   \frac{(2\bk+\bp)_\mu (2\bk+\bp)_\nu q^2 }{ k^4 (\bp+\bk)^2  (q+16   \sigma_0^2 )} \left( \frac{1}{(\bk+\bq)^2} - \frac{1}{q^2}\right) 
  & a_2=2\nn
\Sigma^{(3)}_{\mu\nu} (p) &= -\frac{8}{N} \int_\bq \int_\bk 
   \frac{(2\bk+\bp)_\mu ( 2(\bk+\bq)+\bp)_\nu q^2 }{ k^2 (\bp+\bk)^2 (\bk+\bq)^2 (\bp+\bk+\bq)^2 (q+16   \sigma_0^2 )} 
   & a_3=1\nn
\Sigma^{(4)}_{\mu\nu} (p) &= \frac{16}{N} \int_\bq \int_\bk 
   \frac{(2 \bk+\bp)_\mu (2\bk+\bp)_\nu  }{ k^4 (\bp+\bk)^2(q+32 \sigma_0^2 )}
   \left( \frac{ (2\bk+\bq)_\lambda (2\bk+\bq)_\rho }{  (\bk+\bq)^2 }  -  \frac{ q_\lambda q_\rho }{q^2} \right) 
   \left( \delta_{\lambda\rho} + \frac{q_\lambda q_\rho}{32 q \sigma_0^2} \right) 
   & a_4=2\nn
\Sigma^{(5)}_{\mu\nu} (p) &= \frac{16}{N} \int_\bq \int_\bk 
   \frac{(2 \bk+\bp)_\mu (2 (\bk + \bq) +\bp)_\nu (2 (\bk + \bp)+\bq)_\lambda (2\bk+\bq)_\rho }{ k^2 (\bp+\bk)^2 (\bk+\bp+\bq)^2 (\bk+\bq)^2 (q+32 \sigma_0^2 )}
   \left( \delta_{\lambda\rho} + \frac{q_\lambda q_\rho}{32 q \sigma_0^2} \right) 
  & ~~a_5=1\nn
\Sigma^{(6)}_{\mu\nu} (p) &= -\frac{16}{N} \int_\bq \int_\bk 
  \frac{(2\bk+\bq)_\rho (2\bk+\bp)_\nu }{ k^2 (\bk+\bq)^2 (\bk+\bp)^2 (q+ 32 \sigma_0^2 )} 
  \left( \delta_{\lambda\rho} + \frac{q_\lambda q_\rho}{32 q \sigma_0^2} \right) \delta_{\mu\lambda} 
  + (\mu \leftrightarrow \nu) 
  & a_6 =4 \nn
\Sigma^{(7)}_{\mu\nu} (p) &= \frac{16}{N} \int_\bq \int_\bk 
  \frac{1}{ (\bk+\bq)^2 ( \bk + \bp )^2 ( q + 32 \sigma_0^2) }  
  \left( \delta_{\lambda\rho} + \frac{q_\lambda q_\rho}{32 q \sigma_0^2} \right) \delta_{\mu\lambda} \delta_{\nu\rho}
  & a_7=4 \nn
\Sigma^{(8)}_{\mu\nu} (p) &= \frac{256}{N} \int_\bq \int_\bk 
 \left( \frac{(2\bk+\bp)_\mu (2\bk+2\bp+\bq)_\lambda}{k^2(\bk+\bp)^2(\bk+\bp+\bq)^2} 
  + \frac{(2\bk+\bp)_\mu (2\bk-\bq)_\lambda}  {k^2(\bk+\bp)^2(\bk-\bq)^2} \right) 
  \frac{\sigma_0^2 \delta_{\rho\nu} }{(|\bp+\bq|+16\sigma_0^2)(q+32\sigma_0^2)}  
  \left( \delta_{\lambda\rho} + \frac{q_\lambda q_\rho}{32 q \sigma_0^2} \right)
  & a_8=2 \nn
\Sigma^{(9)}_{\mu\nu} (p) &= -\frac{256\sigma_0^2}{N} \int_\bq \int _\bk
 \frac{1}{k^2(\bk+\bp+\bq)^2}\frac{1}{(|\bp+\bq|+16\sigma_0^2)(q+32\sigma_0^2)}
  \left( \delta_{\mu\nu}+ \frac{q_\mu q_\nu}{32 q \sigma_0^2} \right)
  & a_9=4 \nn
\Sigma^{(10)}_{\mu\nu} (p) &= -\frac{128}{N}  \int_{\bq,\bk,\bl} \Bigg[ 
  \frac{(2\bk+\bp)_\mu (2\bk+\bq)_\lambda }{ k^2 (\bk+\bp)^2 (\bk+\bq)^2} 
  \frac{(2\bl+\bq)_\rho (2\bl+\bp)_\nu }{ l^2 (\bl+\bp)^2 (\bl+\bq)^2} 
  \frac{(\bp-\bq)^2}{(|\bp-\bq|+16\sigma_0^2)( q + 32 \sigma_0^2) }  
  \left( \delta_{\lambda\rho} + \frac{q_\lambda q_\rho}{32 q \sigma_0^2} \right)\nn
  &\hspace{0mm}+\frac{(2\bk+\bp)_\mu (2\bk+\bp+\bq)_\lambda }{ k^2 (\bk+\bp)^2 (\bk+\bq)^2} 
  \frac{(2\bl+\bp+\bq)_\rho (2\bl+\bp)_\nu }{ l^2 (\bl+\bp)^2 (\bl+\bq)^2} 
  \frac{q^2}{(q+16\sigma_0^2)( |\bp-\bq|+ 32 \sigma_0^2) }  
  \left( \delta_{\lambda\rho} + \frac{(\bp-\bq)_\lambda (\bp-\bq)_\rho}{32 |\bp-\bq| \sigma_0^2} \right)
  \Bigg]
   & a_{10}=2 \nn
\Sigma^{(11)}_{\mu\nu} (p) &= \frac{128}{N} \int_{\bq,\bk,\bl}
\Bigg[
 \frac{ \delta_{\mu\lambda}}{k^2(\bk+\bp+\bq)^2} \left( \frac{(2\bl-\bq)_\rho(2\bl+\bp)_\nu}{l^2(\bl-\bq)^2(\bl+\bp)^2} 
  +  \frac{(2\bl+2\bp+\bq)_\rho(2\bl+\bp)_\nu}{l^2(\bl+\bp)^2(\bl+\bp+\bq)^2}\right)
  \times
  \nn
  &
 \hspace{18mm}  \frac{(\bp+\bq)^2}{(q+32\sigma_0^2)(|\bp+\bq|+16\sigma_0^2)}\left( \delta_{\lambda\rho} + \frac{q_\lambda q_\rho}{32 q \sigma_0^2} \right)
 \Bigg]
  & a_{11}=4 \nn
\Sigma^{(12)}_{\mu\nu} (p) &= -\frac{128}{N} \int_{\bq,\bk,\bl}  
  \frac{1}{k^2(\bk+\bp+\bq)^2}\frac{1}{l^2(\bl+\bp+\bq)^2}\frac{(\bp+\bq)^2}{(|\bp+\bq|+16\sigma_0^2)(q+32\sigma_0^2)}
   \left( \delta_{\mu\nu}+ \frac{q_\mu q_\nu}{32 q \sigma_0^2} \right)
  & a_{12}=4 .\nn
  \label{eq:12}
\end{alignat}
These integrals can be evaluated using our algorithm Tensoria as described in the text and in Ref.~\onlinecite{huh13}. 
To extract the critical exponent $\nu$ from the logarithmic singularity and evaluate the finite terms to 
determine the pole position and spectral function,  
we pull out the momentum cutoff ($\Lambda$) dependent part and split the rest into transversal and longitudinal components:   
\beq
\Sigma^{(i)}_{\mu\nu}(p) = \frac{1}{N} \left[ I^{(i)}_T(p)  \left( \delta_{\mu \nu} - \frac{p_\mu p_\nu}{p^2} \right) + I^{(i)}_L(p)  \frac{p_\mu p_\nu}{p^2} \right] + \frac{1}{N}\left[ \frac{14 \Lambda}{3\pi^2} - \frac{96 \sigma_0^2}{ \pi^2}  \log{\frac{\Lambda}{32\sigma_0^2}}  \right ] \delta_{\mu\nu}\;. 
\eeq
Here the term proportional to $\Lambda$ is a consequence of the non-gauge-invariant momentum cutoff, but can be safely absorbed into a shift
in the position of the critical point; such artifacts do not propagate to the universal constants computed here, which are gauge invariant. The logarithmic singularity yields the correlation length $\nu = 1-{48}/{(N \pi^2)}$ described in the main text, and is absorbed after $\sigma_0^2$ is expressed
in terms of $\rho_s$ via Eq.~(\ref{rhos}). 
For the remaining part, we evaluate the analytically continued pole contributions at the location of the $N\rightarrow\infty$ pole position 
($p=-32 \sigma_0^2$). The numerical integration yield to high accuracy (all constants below are in units of $\sigma_0^2$)
\begin{alignat}{3}
I^{(1)}_T  &= -2.882-0.374i 	  &~~ I^{(1)}_L &= -0.213 +0.246i \nn
I^{(2)}_T  &= -0.548-0.265i 	  &~~ I^{(2)}_L &= -0.270 + 0.318i \nn
I^{(3)}_T  &= 0.539+1.197i 	  &~~ I^{(3)}_L &= 0 + 0.098i \nn
I^{(4)}_T  &= 0.476 + 0i		      &~~ I^{(4)}_L &= -0.050+ 0i \nn
I^{(5)}_T  &= 1.895-0.587i     &~~ I^{(5)}_L &= -0.994 + 0i \nn
I^{(6)}_T  &= -0.454 	+ 0i		      &~~ I^{(6)}_L &= 0.548 + 0i \nn
I^{(7)}_T  &= -1.754 	+0i		      &~~ I^{(7)}_L &=-0.274 + 0i \nn
I^{(8)}_T  &= -0.629-0.148i 	   &~~ I^{(8)}_L      &= -0.426 + 0.492i \nn
I^{(9)}_T  &= 2.882+0.374i 	   &~~ I^{(9)}_L       &= 0.213 - 0.246i \nn
I^{(10)}_T  &= -0.606-3.098i   &~~ I^{(10)}_L      &= -0.635 +0.492 i \nn
I^{(11)}_T  &= 1.083+0.148i    &~~ I^{(11)}_L     &= -0.121-0.492 i \nn
I^{(12)}_T  &= -1.128-0.374i   &~~~~ I^{(12)}_L &= 0.061 +0.246i\; . 
\end{alignat}

Combining these, we get 
\bea
I^{tot}_T&=&\sum_i a_i I^{(i)}_T = -9.191 -7.319i \nn
I^{tot}_L&=&\sum_i a_i I^{(i)}_L = -2.906+1.720 i \;. 
\eea

\end{widetext}
\end{document}